\documentclass[preprint,12pt]{aastex}

\newcommand{\as}{\mbox{\arcsec}}

\newcommand{\lsun}{\mbox{L$_\odot$}}

\def\plotfiddle#1#2#3#4#5#6#7{\centering \leavevmode
\vbox to#2{\rule{0pt}{#2}}
\includegraphics{#1}}

\begin{document}

\slugcomment{Accepted for Publication in Astrophysical Journal Letters}

\title{\bf First Science Results From SOFIA/FORCAST: Super-Resolution Imaging of the S140 Cluster at 37\micron}
\author{Paul M. Harvey\altaffilmark{1},
Joseph D. Adams\altaffilmark{2},
Terry L. Herter\altaffilmark{2},
George Gull\altaffilmark{2},
Justin Schoenwald\altaffilmark{2},
Luke D. Keller\altaffilmark{3},
James M. De Buizer\altaffilmark{4},
William Vacca\altaffilmark{4},
William Reach\altaffilmark{4},
and E. E. Becklin\altaffilmark{4,5}
}

\altaffiltext{1}{Astronomy Department, University of Texas at Austin, 1 University Station C1400, Austin, TX 78712-0259;  pmh@astro.as.utexas.edu}
\altaffiltext{2}{Center for Radiophysics and Space Research, Space Science Building, Cornell University, Ithaca, NY 14853;  jdadams@astro.cornell.edu, tlh10@cornell.edu, geg3@cornell.edu, jps10@cornell.edu}
\altaffiltext{3}{Ithaca College Physics Dept., 264 Ctr for Natural Sciences, Ithaca, NY 14850; lkeller@ithaca.edu}
\altaffiltext{4}{SOFIA-University Space Research Association, NASA Ames Reseach Center, Mail Stop N211-3, Moffett Field, CA 94035;  jdebuizer@sofia.usra.edu, wvacca@sofia.usra.edu, wreach@sofia.usra.edu, ebecklin@sofia.usra.edu}
\altaffiltext{5}{University of California Los Angeles, Department of Physics and Astronomy, 405 Hilgard Ave., Los Angeles, CA 90095-1562;  ebecklin@sofia.usra.edu}

\begin{abstract}


We present  37\micron\ imaging of the S140 complex of infrared
sources centered on IRS1 made with the FORCAST camera on SOFIA.  
These observations are the longest wavelength imaging to resolve clearly
the three
main sources seen at shorter wavelengths, IRS 1, 2 and 3, and are nearly at the diffraction limit of
the 2.5-m telescope.  We also obtained a small number of images
at 11 and 31\micron\ that are useful for flux measurement.   Our images cover the
area of several strong sub-mm sources seen in the area -- SMM 1, 2, and 3 -- that are not
coincident with any mid-infrared sources and are not visible in our longer wavelength imaging either.  
Our new observations confirm previous estimates
of the relative dust optical depth and source luminosity for the components in this likely
cluster of early B stars.  We also investigate the use of super-resolution to go beyond
the basic diffraction limit in imaging on SOFIA and find that the van Cittert algorithm,
together with the ``multi-resolution'' technique, provides excellent results.

\end{abstract}

\keywords{infrared: general --- stars: formation --- ISM:individual (S140)}

\section{Introduction}\label{intro}

Sharpless 140 is a relatively diffuse H II region at the edge of the much denser
L1204 molecular cloud that harbors several clusters of young B stars at a distance
of about 900 pc \citep{crampton74}.
The most luminous object in this cloud was first observed in the near-
and mid-infrared by \citet{blair78} and  by
\citet{harvey78} in the far-infrared.
Blair et al. found a very red, compact object with a spectral energy
distribution (SED) similar to that of the BN object in Orion in addition
to several other less remarkable infrared sources.  Using the Kuiper Airborne Observatory, Harvey et al. found
strong, nearly spatially unresolved emission in the far-infrared from
35 - 175\micron.  They fit several crude models to the SED that showed
that the dust optical depth to the central source had to approach unity
in the 50-100\micron\ spectral region and that the total luminosity was equivalent to that
of an early B star.

In the intervening years a host of much higher angular resolution near-infrared and radio
observations have been published.   These have revealed many additional infrared sources likely to be young
stars or protostars as well as energetic outflows similar to those found in many other
molecular clouds harboring young stars, e.g. \citep{evans89, kraemer01, dewit09, preib02,
kurtz99, bally02}.  It also seems likely that one or more of the strong mid-ir sources contains
two or more young B stars \citep{preib02, preib01} and that the radio jet from IRS1 is likely
to be due to an ionized equatorial wind \citep{hoare06}.  Very cold, dense condensations only emitting
longward of $\lambda \sim$ 300\micron\ are also seen and may be even younger
stages of star formation within the molecular cloud \citep{minchin95}.

Most of the luminosity of very young stellar objects and protostars is emitted in the
far-infrared, wavelengths that are essentially invisible to ground-based telescopes due
to absorption by water vapor.  Until the recent commissioning of the Herschel Space
Observatory (Pilbratt et al. 2010) and the Stratospheric Observatory for Infrared Astronomy (SOFIA; Gehrz et al. 2011),
the far-infrared, however, has suffered from a substantial deficit in angular resolution
compared to neighboring wavelengths observable with large ground-based telescopes.
High mass young stars typically form in dense clusters, and this puts a special premium on the
value of high angular resolution to resolve individual components to understand their
evolutionary state.

We present here the first imaging observations of the S140 cluster at wavelengths longward of
the ground-based atmospheric window at $\lambda \sim$ 20 -- 25\micron\ with an angular resolution 
of a few arcseconds.  With this resolution we
are able to distinguish the three main luminosity sources seen at shorter wavelengths
as well as diffuse emission between the objects.  We describe the observations in \S\ref{obs}
and then in \S\ref{super} describe some efforts to improve the angular resolution with
various deconvolution techniques.  In \S\ref{compare} we compare our 31 and 37\micron\ maps
with those at shorter and longer wavelengths and show the spectral energy distributions 
for the individual members of this cluster,  and in \S\ref{summ}  we summarize our
results.

\section{Observations and Data Reduction}\label{obs}

Our observations were made with the FORCAST camera (Herter et al. 2012) on board SOFIA 
on the night of UT 1 Dec 2010.
The FORCAST camera has two separate 256$\times$256 infrared arrays that cover the wavelength range from
$\sim$ 5 -- 40\micron\ in a variety of bands with 0.768$\times$0.768\as\ pixels.  The two arrays 
can observe simultaneously
through a dichroic beamsplitter that divides the wavelength range shortward and longward of 26\micron,
or the long wavelength array can be used by itself without the intervening dichroic. 
The observations presented here were made with the 37.1\micron\ filter (hereafter, 37\micron) without the beamsplitter
and at 11.1 and 31.5\micron\ with the beamsplitter (hereafter ``11'' and ``31''\micron).

In addition to the observations of S140, a number of other science investigations were undertaken
that are also reported in this issue, and several calibration targets were observed for both flux and
point-source-response (PSF) calibration.  
The FORCAST images were calibrated to a flux density per pixel using a standard instrument
response derived from measurements of standard stars
and solar system objects over several 
flights (Herter et al. 2012). The color correction from a flat spectrum source to that of
S140 is negligible ($< 1\%$). The estimated $3\sigma$ uncertainty in the
calibration due to variations in flat field, water vapor burden, and altitude
is approximately $\pm 20\%$.
The point source response has been derived from averages over several flights as described in detail
by Buizer et al. (2012).  We discuss below the relatively minor uncertainties involved in using
this average PSF, and tests of its applicability using $\mu$ Cep data taken on the same flight as S140.
The nominal diffraction limit of SOFIA at 37\micron, 1.019$\lambda/$D with its 14\% central
obscuration, is 3.1\as, and our observations
show that achieved resolution is quite close to this limit, $\sim$ 4\as\ FWHM.  At shorter wavelengths the SOFIA PSF
is affected more by telescope pointing jitter and seeing, with seeing being dominant shortward of 5\micron\ because
of the shear layer just above the telescope \citep{seeing00}.

Our images of both S140 and $\mu$ Cep were obtained with standard thermal infrared
techniques of chopping and nodding.  
Because the cluster is relatively compact, we were able
to chop and nod completely on the $\sim$ 200 arcsec square array.  
The chop frequency used was 2 Hz, and chop amplitude of
90\as\ with a nod amplitude also of 90\as\ perpendicular to the chop direction.
Each observation resulted
in 120 seconds of on-chip integration as the result of one ABBA nod pair.
We obtained 10 useful observations of S140 and two of $\mu$ Cep at 37\micron,  and two of S140 
at 11 and 31\micron.  The individual
observations were co-added by cross-correlating all the subsequent images with the first.
The sky rotation was essentially negligible between the individual S140 observations.
The telescope software to enable accurate absolute pointing was
not yet operational for these observations, so we used the highest resolution, nearest wavelength
observations (24.5\micron) from a large ground-based telescope \citep{dewit09} to 
establish the absolute coordinate system
for our observations.  For the purposes of deconvolution discussed later, the S140 frames were left in the
original instrument coordinate system to match those of the PSF calibration image.

\begin{figure}
\plotfiddle{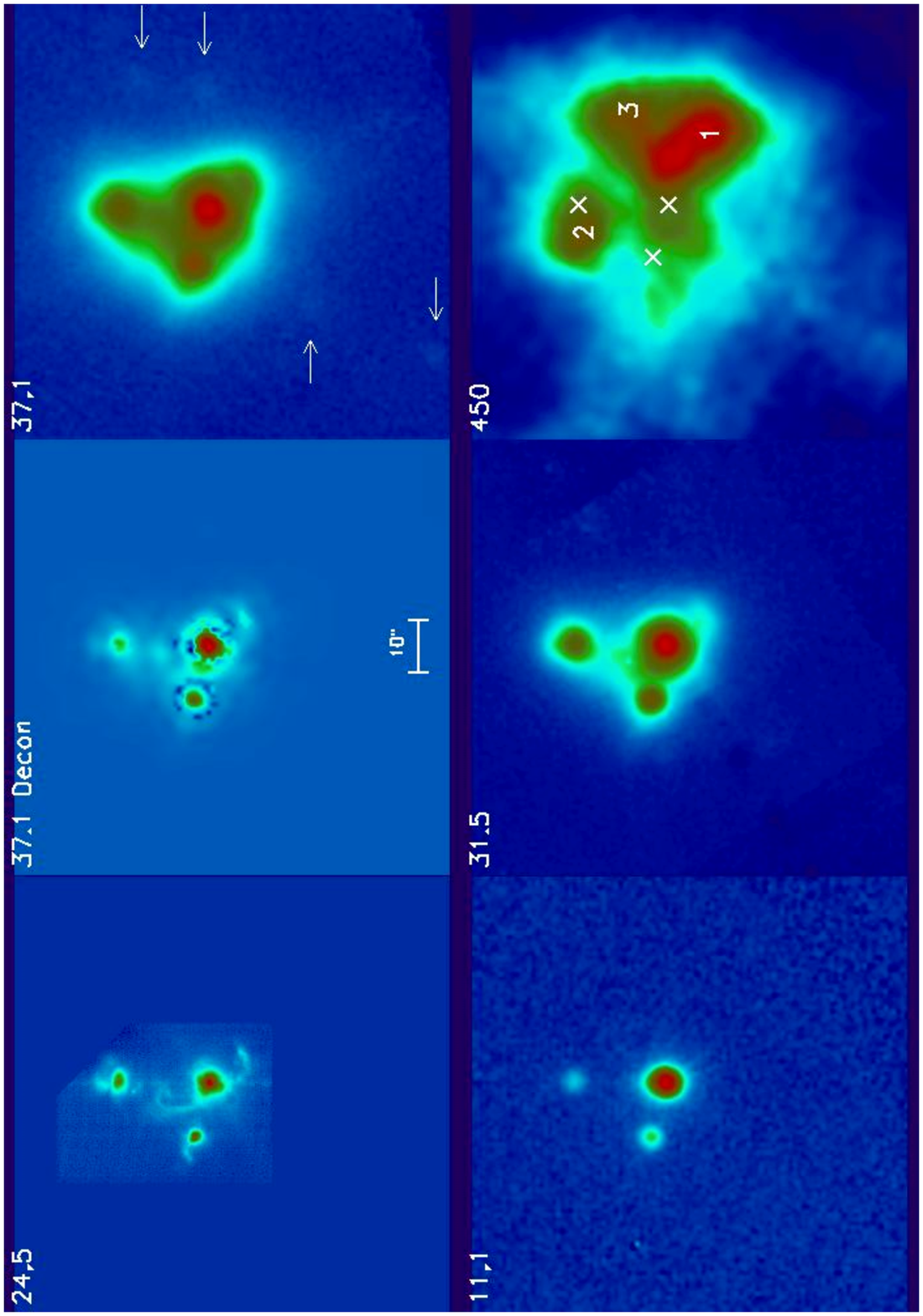}{7.0in}{-90}{70}{70}{-250}{500}
\figcaption{\label{decons140}
Top row (left to right) - 24.5\micron\ Subaru image of \citet{dewit09}, deconvolved 37\micron\ image from this
paper, and observed 37\micron\ image from this paper; Bottom row (left to right) - observed 11\micron\ and 31\micron\
images from this paper, and 450\micron\ images from the SCUBA archive (http://cadcwww.dao.nrc.ca/jcmt/search/product) with the positions of IRS 1,2, and 3 marked with ``X'', as well as the positions of the sub-millimeter 
peaks SMM1--3.  North is up and east to the left. The locations of 4 faint sources that are
coincident with {\it Spitzer} IRAC sources are indicated by arrows in the upper right panel.}
\end{figure}

Figure \ref{decons140} shows the results of the processing described so far for S140 as well as related
images from the literature discussed below.
Using aperture photometry we have derived flux densities for the three obvious components of the image
as well as for two additional local flux maxima.  These results are listed later in Table \ref{apflux} for a
9" diameter aperture.  We discuss these further in \S\ref{compare} after we describe our
deconvolution work and the resulting higher resolution image.  
The statistical uncertainties in these fluxes are negligible because of the very high signal-to-noise
ratio, well over 100, for all the sources except for the faintest objects at 11\micron\ where S/N $\sim$ 10.  
Table \ref{apflux} also lists the luminosities
of these ``objects'' for the assumed distance of 960 pc, using the 50--100\micron\ flux densities of
\citet{lester86}.

\section{Deconvolution and Super-Resolution}\label{super}

With NASA's previous airborne observatory, the KAO, \citet{lester86} for example
showed that significant improvements in the apparent angular resolution in 
far-infrared images were possible through super-resolution techniques such as
maximum entropy deconvolution.  With the KAO's instrumentation these techniques
were limited to 1-D scans rather than 2-D images.  In the intervening years, the
improvement in detectors and in analysis techniques suggests the possibility
of substantial resolution improvements in 2-D far-infrared images with SOFIA.
To that end we have experimented with a variety of algorithms and parameters
available as part of the ``MRA'' package \citep{starck02, pantin96} to improve
the resolution in our S140 image.  Our best results were obtained using the van-Cittert
algorithm with multi-resolution support for regularization.  With this mode the
final deconvolved images did not depend strongly on the number of iterations, and therefore
this was used for all our results below.  For example, the final image shown used 500 iterations,
but the image changed very modestly after as few as 20 iterations.  Additionally, we did not force positivity
in the image.

\begin{figure}
\plotfiddle{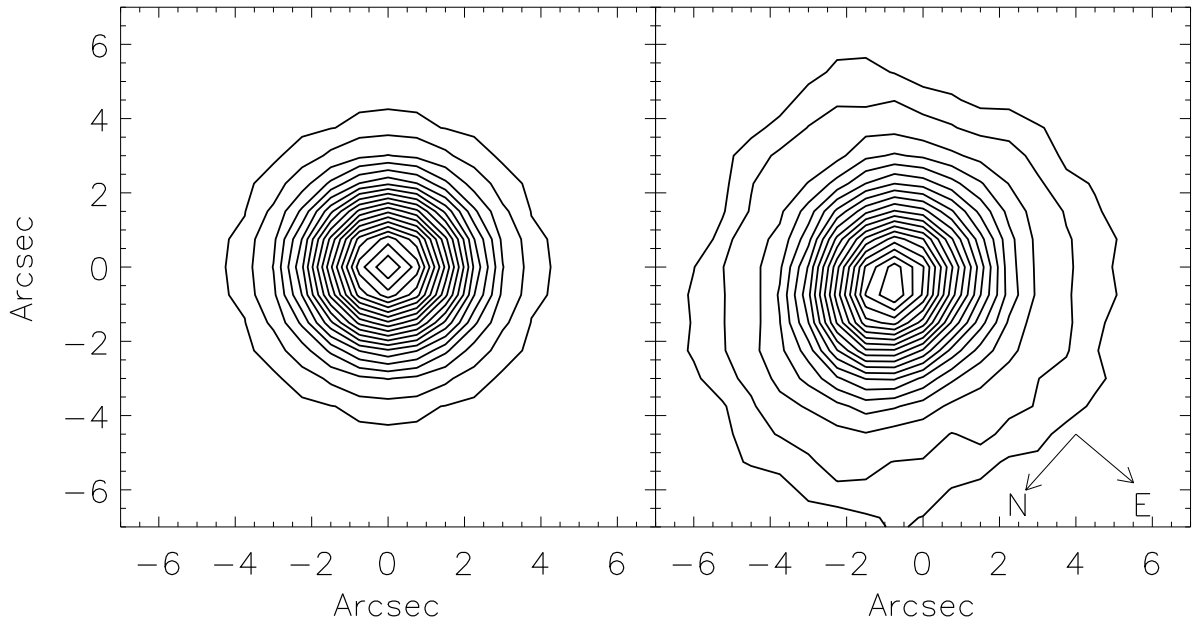}{7.0in}{0}{80}{80}{-250}{-10}
\figcaption{\label{psffig}
Left - Derived PSF as described in the text.  Contours are at levels of 2.5, 5, 10, 15,..., 90, and 95\% of
the peak.  Right - Observed image of the nominal PSF calibrator, $\mu$ Cep, with contours at the same level.}
\end{figure}

As mentioned above we have used
the PSF derived by  Buizer et al. (2012) and done several tests to give us confidence in its
applicability to the S140 images.
First we deconvolved the observed $\mu$ Cep image and compared the result with the
much higher angular resolution, but shorter wavelength 24.5\micron\ image of \citet{dewit08}.
We found the spatial extent and spatial asymmetry to be very similar.  We also found the amplitude
of the wings to the northeast and southwest to be relatively larger at 37\micron, which is entirely
consistent with the expected dust temperature of order 150K at a radius of a few arcseconds.
In a second test of our derived PSF, we produced several similar PSF's with small changes
in the width in both telescope axes, elevation and cross elevation, and then deconvolved the
S140 image with these modified PSF's.  In all cases the major details of the deconvolved images
were essentially the same as for the derived PSF.  So we are confident that none of our conclusions depends on
the small uncertainties in our knowledge of the PSF.
Finally, Figure \ref{decons140} shows the observed and deconvolved images of S140 at 37\micron\ together with the
observed Subaru image \citep{dewit09}.  It is clear that virtually all the structure seen at the
0.6\arcsec\ diffraction limit of Subaru at 24.5\micron\ is also visible at 37\micron\ in the
deconvolved image, albeit at somewhat lower resolution.  With the native 37\micron\ resolution of
$\sim$ 4\arcsec, however, much of this structure is smoothed to the point of being barely detectable.
The final derived PSF is equivalent to the theoretical diffraction-limited PSF convolved with
a Gaussian of roughly 1.5\as\ FWHM, and the apparent resolution of the deconvolved S140 image is $\sim$ 2.4\arcsec.
Figure \ref{psffig} shows the derived
PSF along with the observed $\mu$ Cep image illustrating the larger extent of $\mu$ Cep at 37\micron.

\section{Comparison With Ground-Based Mid-Infrared and Radio Observations}\label{compare}

We have already noted above the excellent agreement between our 37\micron\ deconvolved image and the
much higher resolution 24.5\micron\ image of \citet{dewit09}.  In particular, in addition to the
good agreement in position of the three main peaks, the agreement in morphology of the low level,
diffuse emission north of IRS 1 and also to the southwest suggests that the deconvolution
process has not developed spurious features as might have been the case with a poor
estimate for the PSF or a poor algorithm.   There are probably still some residual
positional problems in our 37\micron\ image at the level of $\sim$ 1\arcsec\ due to the inherent 
uncertainties in fixing the aboslute positions by registering to the shorter wavelength ground-based
images,
but the agreement between the two images is remarkably satisfying.  As a corollary it is also clear
that there are no new sources present in the 37\micron\ data, either raw or deconvolved, that are not
present in the Subaru 24.5\micron\ image.  Most importantly, we do not see any evidence for emission above
the background from either of SMM 1, SMM 2, or SMM 3, the 450\micron\ sources 
discovered by \citet{minchin95}.  At 37\micron\ any emission from these sources must be below a
level of $\sim$ 30 Jy to be invisible in our image.
This confirms their likely interpretation as very cold, prestellar condensations.  Four very faint sources
in the 37\micron\ image are indicated in Figure \ref{decons140} that have counterparts in the {\it Spitzer}
IRAC images in the {\it Spitzer} archive, presumably low luminosity young stellar objects.

\begin{figure}
\plotfiddle{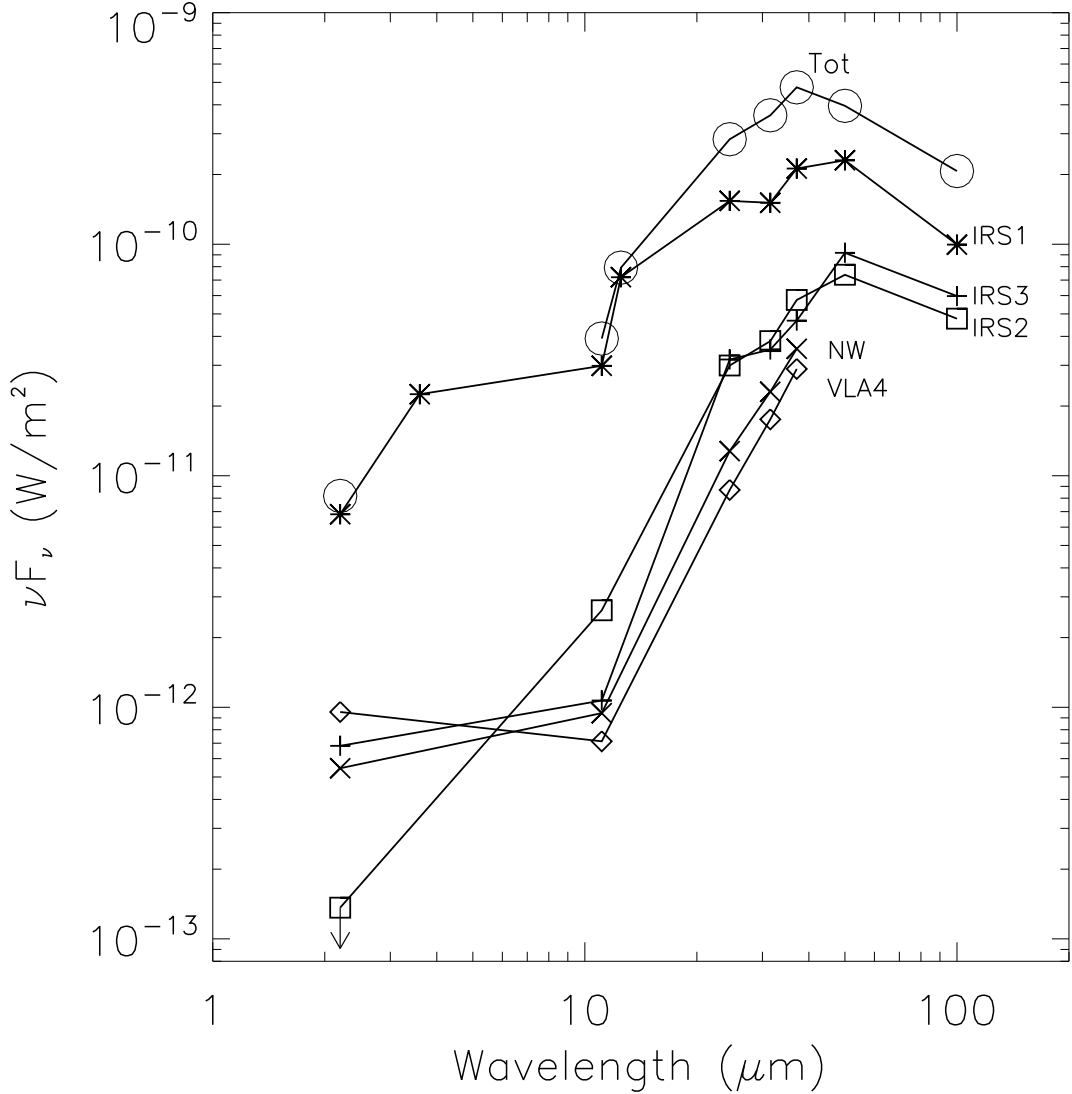}{7.0in}{00}{80}{80}{-250}{-10}
\figcaption{\label{sed}
Spectral energy distributions of the various sources detected in the S140 region
at 37\micron.  For the 37\micron\ flux densities we have used the 9\arcsec\ aperture
fluxes listed in Table \ref{apflux} from the deconvolved map.  For the 11 and 31\micron\ fluxes from this
study, we used the 9\arcsec\ aperture fluxes in Table \ref{apflux} from the observed images.  At the shorter wavelengths
we have used the 9\arcsec\ aperture fluxes from \citet{evans89} at 1--2.5\micron, the
3.45\micron\ flux for IRS1 from \citet{blair78}, and measured 9\arcsec\ aperture fluxes
from the 24.5\micron\ map of \citet{dewit09}.  At 50 and 100\micron\ we used the fluxes
estimated by \citet{lester86} for the three major sources.}
\end{figure}

\begin{figure}
\plotfiddle{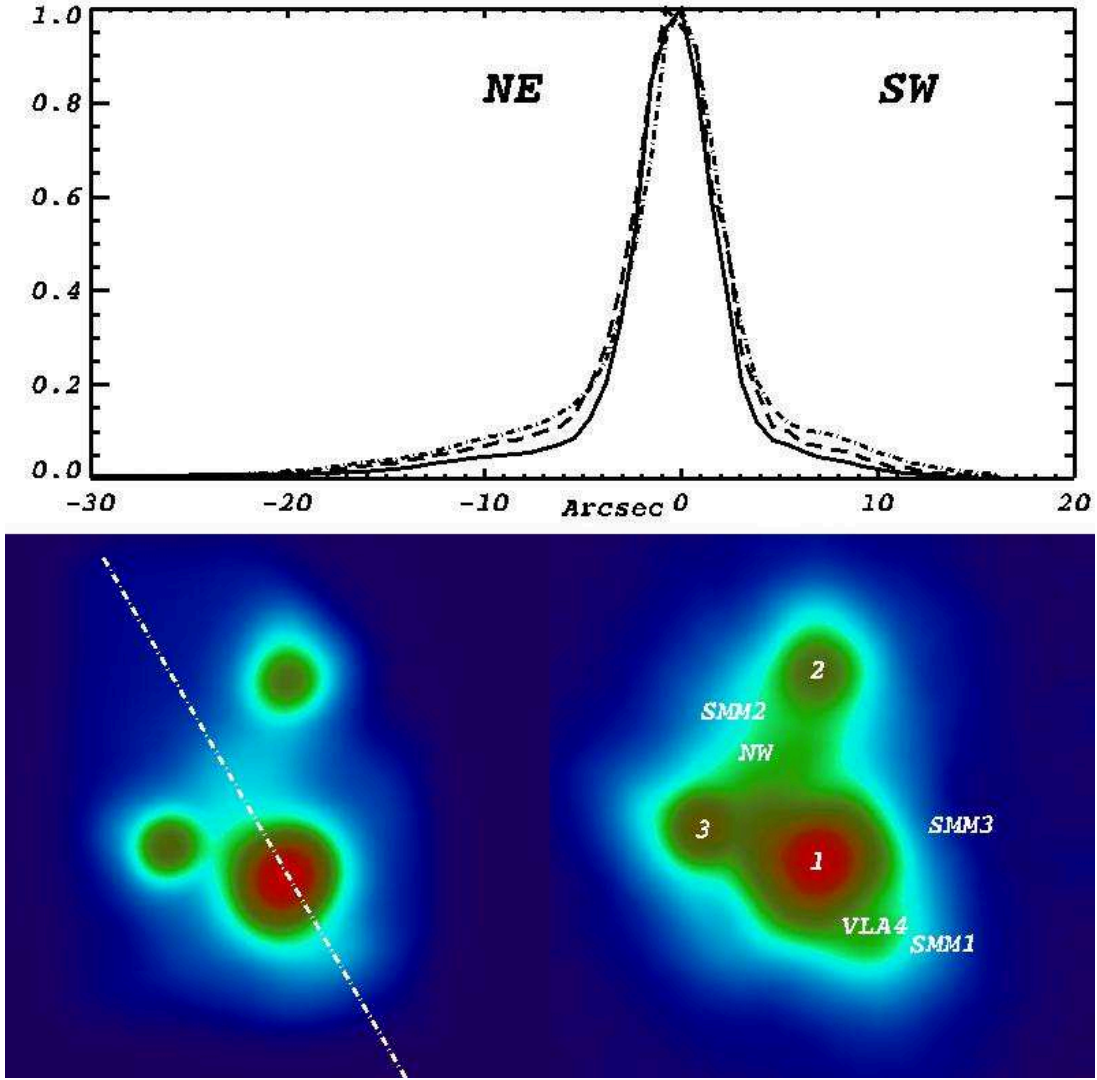}{6.5in}{90}{70}{70}{250}{40}
\figcaption{\label{recon}
Bottom Panel - 24.5\micron\ image of \citet{dewit09} convolved with 37\micron\ PSF (left), and our observed 37\micron\
image illustrating the more extended, diffuse emission at the longer wavelength.  The diagonal line
through IRS 1 marks the 1-D cut along which the relative fluxes are shown in the upper panel,
showing quantitatively
the higher level of extended emission at the longer wavelengths. The approximate positions of
all the components of this source mentioned in the text are indicated in the lower right panel.
Upper Panel - relative fluxes along the diagonal line below for the observed 37\micron\ image (dash-dot),
the observed 31\micron\ imaged convolved to the 37\micron\ resolution (dash), and the 24.5\micron\
image of \citet{dewit09} convolved to the 37\micron\ resolution (solid).}
\end{figure}

As noted earlier, Table \ref{apflux} lists flux densities derived from our 37\micron\ image.  We list both
the values measured using the raw image, as well as from the deconvolved image.  Since the deconvolution
algorithm conserves flux, the total flux densities for both methods are identical, but we believe the
deconvolved image provides a more realistic measurement of the flux densities in small beams required
for a reasonable comparison of the relative fluxes and colors of the individual components in this region.
Table \ref{apflux} also lists the observed 11 and 31\micron\ flux densities for a 9\arcsec\ aperture.
Figure \ref{sed} shows the spectral energy distributions for the various local maxima that have been
identified previously in this region and which are clearly detected in our deconvolved map.  This shows
graphically the fact, as noted by previous investigators, that the coldest mid-infrared condensations in this region
are the diffuse local flux maxima called ``NW'' and ``VLA 4'' (labelled in Figure \ref{recon}), while the hottest SED belongs to IRS 1.  IRS 2 and 3 exhibit
SED's that are intermediate in color temperature with IRS 3 being the cooler source.
Quantitatively, IRS 1 dominates the shorter wavelength emission with 75\% of the total emission at
11\micron, but at 37\micron\ it contributes less than 40\% with the rest arising from the cooler sources,
IRS 2, 3, NW, VLA4, and the surrounding diffuse emission.

In their original KAO study of S140 \citet{harvey78} showed that the dust optical depth was likely
close to unity in the far-infrared based on the resolving power possible with that instrumentation.
With the substantially higher spatial resolution available with SOFIA/FORCAST we can re-examine
this conclusion.  For example, the ratio of 37\micron\ to 53\micron\ flux densities shown in
Figure \ref{sed} for IRS1 implies a color temperature of order 85K.  If we assume that half of the total
flux arises in the central 4\as\ core, this implies $\tau \sim$ 0.5 at 53\micron, quite consistent
with the original KAO study.  

In order to compare quantitatively the diffuse emission in our 37\micron\ map with that seen
at much higher angular resolution at 24.5\micron\ by \citet{dewit09}, it is necessary to
compare the two images at identical levels of resolution.  This technique was described as 
``beam matching'' by \citet{lester86}.  Since the native resolution at 37\micron\ is more than
a factor of 10 worse than that at 24.5\micron\, it is sufficiently accurate to simply convolve
the 24.5\micron\ image with the 37\micron\ PSF.
Figure \ref{recon} shows the original, undeconvolved 37\micron\ SOFIA image together with the 24.5\micron\
image of \citet{dewit09} smoothed to what would have been observed with our 37\micron\ SOFIA PSF.  This
shows that there
is substantially less extended diffuse emission at 24.5\micron.  Figure \ref{recon} also shows
the relative 24.5, 31, and 37\micron\ fluxes along a one dimensional cut through IRS 1 from northeast 
to southwest along the line
shown in the figure.  This comparison shows quantitatively the higher level of extended, diffuse
emission at the longer wavelengths, consistent with a radial temperature gradient away from the center
of heating.  This effect can also be seen in the more steeply rising 24.5--37\micron\ SED slopes noted earlier
for the sources labeled ``NW'' and ``VLA4'' well away from IRS1.

In order to fully model the various components of this compact cluster, it will be necessary
to obtain spatially resolved images at the peaks of their SED's, i.e $\lambda \sim$ 70\micron.
For the 3 brightest sources, this will be possible with SOFIA using the upcoming HAWC far-ir
camera.  To understand the structure of the cores of the sub-millimeter sources would, however,
require the sensitivity and higher angular resolution of {\it Herschel} with its PACS and
SPIRE cameras.

\section{Summary}\label{summ}

We have demonstrated the utility of the FORCAST camera on SOFIA for mapping compact regions of star formation at wavelengths
beyond the familiar ground-based mid-IR atmospheric windows.  This allows us to define the SED's
and colors of the major emission sources in S140 with unprecedented angular resolution in the far-IR.
Since the data were taken with quite high S/N, we were able to successfully pursue spatial deconvolution
of the image and show that the basic structure which appears at $\lambda \sim$ 25\micron\ with sub-arcsecond
resolution from large ground-based telescopes can also be identified in our far-IR images.  
We find emission from {\it all} the major sources seen in this cluster at shorter wavelengths from ground-based
telescopes and confirm that the coolest sources are IRS3 and the diffuse knots, VLA4 and NW.  
We also see
no evidence for emission from the sub-mm peaks in this region.

\section{Acknowledgments}

This work is based on observations made with the NASA/DLR Stratospheric Observatory for Infrared Astronomy (SOFIA). SOFIA science mission operations are conducted jointly by the Universities Space Research Association, Inc. (USRA), under NASA contract NAS2-97001, and the Deutsches SOFIA Institut (DSI) under DLR contract 50 OK 0901. Financial support for FORCAST was provided to Cornell by NASA through award 8500-98-014 issued by USRA.  Support at the
University of Texas was provided by USRA award 08521-012..
We also especially thank W. de Wit for providing the digital versions of the 24.5\micron\ images for
both $\mu$ Cep and S140 and acknowledge insightful comments from D. Lester and an anonymous referee..
This research used the facilities of the Canadian Astronomy Data Centre operated by the National Research Council of Canada with the support of the Canadian Space Agency.

\clearpage

\begin{deluxetable}{lcccccc}
\tablecolumns{7}
\tablecaption{Aperture\tablenotemark{a} Flux Densities\tablenotemark{c} (Jy) of Sources in S140\label{apflux}}
\tablewidth{0pc}
\tablehead{
\colhead{Image}    &
\colhead{IRS 1}    &
\colhead{IRS 2}  &
\colhead{IRS 3}   &
\colhead{NW} &
\colhead{VLA 4}   &
\colhead{Total\tablenotemark{b}}
}

\startdata

Observed 37\micron\ & 2176   &  485  & 669 &  465 & 381 & 6730 \\
Deconvolved 37\micron\ & 2624   &  577  & 710 &  438  & 358 & 6730 \\
Observed 11\micron\ & 110   &  4.0  & 9.7  &  3.5 & 2.6 & 145 \\
Observed 31\micron\ & 1585   &  368  & 401 &  242 & 184 & 3780 \\
Luminosity (\lsun) & 10,700 & 3200 & 2900 & 800 & 650 & 19,150 \\

\enddata
\tablenotetext{a}{For 9\arcsec\ diameter aperture.}
\tablenotetext{b}{Summed over entire map shown in Figure \ref{decons140}.}
\tablenotetext{c}{Absolute calibration uncertainty = 20\%; negligible statistical uncertainty.}

\end{deluxetable}

\clearpage

\clearpage

\end{document}